# Distibuted Massive MIMO Channel Estimation and Channel Database Assistance

Arkady Molev-Shteiman
*Radio Algorithm Research*
Huawei Technologies
Bridgewater, NJ USA
arkady.molev.shteiman@huawei.com

Laurence Mailaender
*Radio Algorithm Research*
Huawei Technologies
Bridgewater, NJ USA
laurence.mailaender@huawei.com

Xiao-Feng Qi
*Radio Algorithm Research*
Huawei Technologies
Bridgewater NJ USA
xiao.feng.qi@huawei.com

*Abstract*—Due to the low per-antenna SNR and high signaling overhead, channel estimation is a major bottleneck in Massive MIMO systems. Spatial constraints can improve estimation performance by exploiting sparsity. Solutions exist for far field - beam domain channel estimation based on angle of arrival estimation. However, there is no equivalent solution for near field and distributed MIMO spatial channel estimation. We present a solution- source domain channel estimation- that is based on source location estimation. We extend this to employ a 'Channel Database' incorporating information about the physical scattering environment into channel estimation. We present methods for generation, storage and usage of the Channel Database to assist localization and communication.

## I. INTRODUCTION

As the number of antennas increases in modern Massive MIMO communications, channel estimation becomes increasingly burdensome. In fact, channel estimation may dominate the receiver algorithmic complexity, impose large signaling overhead, and become the prime limiting factor in the overall system capacity [1].

Estimation error generally increases when more uncorrelated parameters have to be estimated from the same data record. This is a serious performance issue for Massive MIMO where $N$ may be 100, 1000, or more. It is commonly assumed that the narrowband channel between an $M$ element array and a single antenna user is characterized by $M$ independent parameters. A step forward is achieved if the channel can instead be modeled as a sum of $K$ directional beams ($K<<M$). In such cases the number of parameters to be estimated is dramatically reduced to $4K$ (assuming an angle, delay, and complex amplitude for each beam). Such beam-domain techniques have been shown to offer powerful performance gains [2] [3]. Previous works have assumed the far-field scenario for spatial channel estimation, which cannot be applied to near-field sources or distributed antennas where the beam concept doesn't work. We generalize this by introducing a spatial channel estimation algorithm for near-field that represents the channel as a sum of sources rather than beams, therefore we estimate source locations rather than angles of arrival. This near-field approach is a more general and convenient representation for distributed Massive MIMO systems.

Localization by Direct Positioning in [4] used virtual transmitters to model the multipath and thereby improve location accuracy. Recently, this was extended to estimate the virtual positions simultaneously with the mobile's position [5], which does not require any prior database information, although at the cost of high complexity. Location assisted communication was proposed in [6], however they concentrated mainly on higher layer issues and not channel estimation. Authors in [7] [8] [9] present methods where location information reduces the search space to make channel estimation faster and less complex.

Taking the location concept even further, it is possible to claim that if the radio environment were known exactly, the channel becomes completely deterministic given the user and array locations, meaning there are in fact only 5 parameters to estimate (the user's $x,y,z$ position and complex amplitude) regardless of the number of radio antennas. In other words, the user's position provides a strong 'ray tracing' constraint. This concept holds the potential for great improvements in capacity, if the user location and the radio environment (walls, buildings, etc.) can be known to sufficient precision.

Channel databases are used in channel change prediction, scheduling, and other algorithms for system robustness [10] [11]. A popular approach to these goals is using an RF fingerprinting database [12]. Our paper takes a database approach, but our emphasis is on channel estimation taking the deterministic ray tracing model of the environment into account. This is a highly 'compressed' form of environmental information that improves both location and channel estimation.

## II. MULTISOURCE CHANNEL ESTIMATION

### A. Channel Representation

The concept of modeling multipath as originating from an additional 'virtual source' is well known in the engineering literature, for example, to model the ground bounce in mobile radio channels [13].

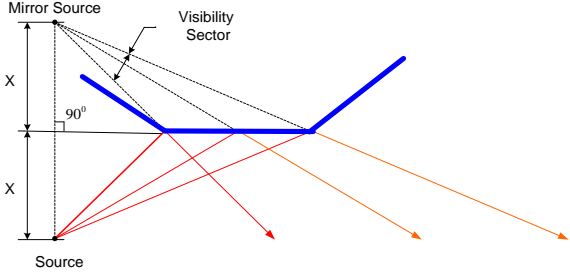

Figure 1: Reflection as Mirror Source

This single bounce idea is easily extended to more complex propagation scenarios where each wall, building, etc. requires an additional virtual (or, 'mirror') source to model its contribution to the received signal. Multiple bounces can also be handled in a like manner by adding more mirror sources. We will call a source that produces $K$ reflections a source of order $K$. In Figure 2 we see an example of ray tracing in a hallway with two walls. If we neglect higher than second order sources, then this environment includes the following sources:

- Original Main Source
- Virtual Mirror Source 1 – The reflection of the main source from wall 1. It is a first order reflection.
- Virtual Mirror Source 2 - The reflection of main source from wall 2. It is a first order reflection.
- Virtual Mirror Source 2.1 - The reflection of mirror source 2 from wall 2. It is a second order reflection.

But the AP sees only three sources: main source, mirror source 1 and mirror source 2.1. Mirror source 2 is not included as the AP is outside the visibility sector of Mirror Source 2.

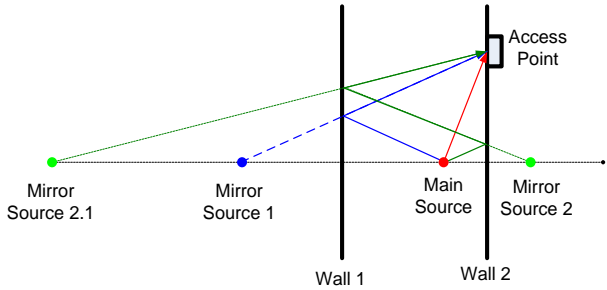

Figure 2: Multisource Ray Tracing

The multisource channel is represented as the superposition of $K$ real and mirror sources [14] and may be expressed:

$$h_m = \sum_{k=0}^{K-1} g_k \cdot Str(l_k, l_m) \quad (1.1)$$

Where: $m$ is the AP antenna index

$M$ is the number of AP antennas

$l_m = (x_m, y_m, z_m)$ is the AP antenna $m$ location

$l_k = (x_k, y_k, z_k)$ is the source $k$ location

$g_k$ is the source $k$ complex amplitude

The function $Str(l_k, l_m)$ represents the line of sight channel between point with location $l_m$ and point with location $l_k$. Here we use the 'near-field' description in which angles do not appear. If the steering vector includes both the near-field phase response and the free-space path loss, is it written as,

$$Str(l_k, l_m) = \frac{\exp(j \cdot 2 \cdot \pi \cdot |l_k - l_m|/\lambda)}{(4 \cdot \pi \cdot |l_k - l_m|/\lambda)^2} \quad (1.2)$$

The notation $|l_k - l_m|$ represents distance between two points:

$$|l_k - l_m| = \sqrt{(x_m - x_k)^2 + (y_m - y_k)^2 + (z_m - z_k)^2} \quad (1.3)$$

The source amplitude $g_k$ is a product of $N$ gains $r_{k,n}$ of its all reflections and transmitter complex amplitude $g_{TX}$.

$$g_k = g_{TX} \cdot \prod_{n=0}^{N-1} r_{k,n} \quad (1.4)$$

Channel estimation can be performed by estimating the $K$ source spatial parameters $(\hat{g}_k, \hat{l}_k)$, and applying:

$$\hat{h}_m = \sum_{k=0}^{K-1} \hat{g}_k \cdot Str(\hat{l}_k, l_m) \quad (1.5)$$

B. Channel Estimation Algorithm

There are many ways to estimate the source location (MUSIC, correlation, etc.) Here we propose a correlation-based method due to its low complexity (a MUSIC-based approach is taken in our companion paper [15]).

The Channel Estimation has following steps:

- Antenna domain channel estimation:

$\bar{h}_m$ for $m = 0, 1, ... M - 1$

- Find list of correlation peak (virtual sources locations):

$$\hat{l}_k = \arg\max_l \left( \left| \sum_{m=0}^{M-1} \bar{h}_m \cdot (Str(l, l_m))^* \right| \right) \text{ for } k = 0, 1, ... K - 1 \quad (1.6)$$

The notation $(\ )^*$ represents the conjugate operation.

The search starts with a coarse resolution, $\lambda/4$ and then zooms in with more precise resolution. In our simulation

we did a few iterations of zooming until we got to a resolution of $\lambda/64$

- Given the estimated source locations we estimate the source complex amplitude according to:

$$\hat{g}_k = \frac{1}{M} \cdot \sum_{m=0}^{M-1} \left( \bar{h}_m \cdot \left( Str(\hat{l}_k, l_m)^* \right) \right) \quad \text{for } k = 0,1,...K-1 \quad (1.7)$$

- From the spatial parameters $\left(\hat{g}_k, \hat{l}_k\right)$ we reconstruct the channel according to (1.5).

*C. Simulation*

We simulate the signal propagation in a 2D room with dimensions $W = 6.4m$ by $D = 6.4m$. The carrier frequency is 1.5GHz – the wavelength is 0.2m. We assume only first order reflections, so there are 5 sources: 1 original and 4 virtual. We assume that the wall reflection coefficient is equal to 1. The simulation setup is shown in figure below:

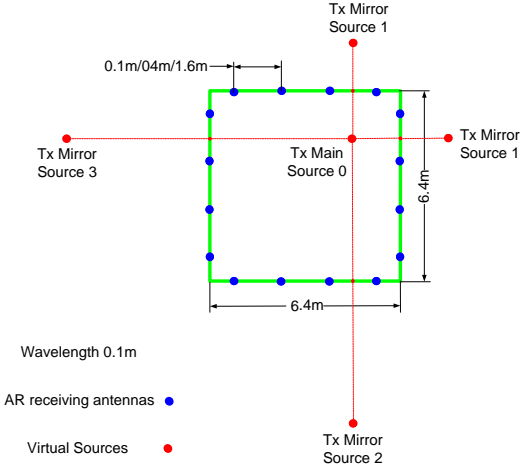

Figure 3: Simulation set up

The UE location is uniformly distributed over the room.

The number of AP antennas depends on spacing between them. We simulate following scenarios:

| Spacing between Antennas | $0.5 \cdot \lambda$ | $2 \cdot \lambda$ | $8 \cdot \lambda$ |
|---|---|---|---|
| Number of Antennas | 256 | 64 | 4 |

Table 1: Simulation Scenario

In order to capture the original and all first-order reflections we search for peak over the interval from *–W* to *2W* and from *–D* to *2D*.

Simulation results for different numbers of antennas are shown in Figure 4.

Because we use antenna domain channel estimation as an input, we present the Error Vector Magnitude (EVM) of the multisource domain channel estimation as a function of antenna domain channel estimation EVM.

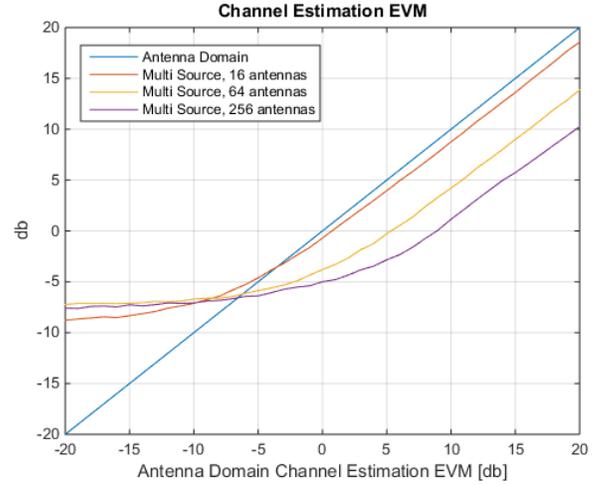

Figure 4: Source domain channel estimation results

### III. MULTISINK CHANNEL ESTIMATION AND CHANNEL DATABASE ASSISTANCE

*A. Channel Representation*

Previous authors have used the Virtual Source approach to model multipath. We believe we are the first to extend this to Virtual Sinks. Due to channel reciprocity we may represent the channel as the superposition of virtual receivers and as well as a superposition of virtual transmitters. An example of this duality is shown in

Figure **5**.

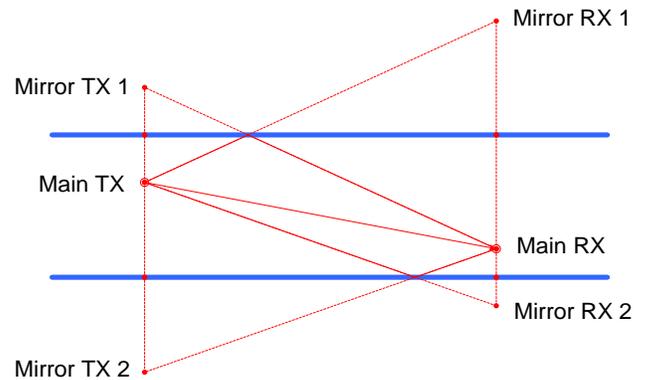

Figure 5: Multi Source/Sink Channel representation duality

In the same manner as the multisource channel was expressed by (1.1) we may express the multisink channel as:

$$h_m = g_{TX} \cdot \sum_{k=0}^{K-1} g_{m,k} \cdot Str(l_{TX}, l_{m,k}) \quad (1.7)$$

Where: $l_{TX}$ is the user transmitter location.

$g_{TX}$ is the user complex amplitude.

$l_{m,k}$ is the AP antenna $m$, virtual sink $k$ location

$g_{m,k}$ is the AP antenna $m$, virtual sink $k$ amplitude

The virtual sink complex amplitude is a product of all reflections and is given by:

$$g_{m,k} = \prod_{n=0}^{N-1} r_{m,k,n} \quad (1.8)$$

where $r_{m,k,n}$ is the reflection coefficient number $n$ of sink $k$ of received antenna $m$.

In reality, due to blockage and limited size of the reflection surface of each virtual sink, there is a 'visibility sector' that defines its reception zone. The nature of the visibility sector is shown in Figure 1. The multisink channel model that takes account of the sink's visibility sectors is given by:

$$h_m = g_{TX} \cdot \sum_{k=0}^{K-1} g_{m,k} \cdot Cvs(l_{TX}, s_{m,k}) \cdot Str(l_{TX}, l_{m,k}) \quad (1.9)$$

Where $Cvs(l, s_{m,k})$ is the indicator function that checks whether location $l$ belongs to visibility sector $s_{m,k}$.

$$Cvs(l_{TX}, s_{m,k}) = \begin{cases} 1 & \text{if } (l_{TX} \in s_{m,k}) \\ 0 & \text{else} \end{cases} \quad (1.10)$$

Expression (1.9) uses the vector of $M \times K$ sink locations, sectors and amplitudes $(g_{m,k}, s_{m,k}, l_{m,k})$ which encodes the local wall positions, and is independent of UE location so the same vector applies to any user. Once the AP is installed this vector can be pre-computed and stored in a database, as the surrounding environment will be quite stable. We call it a Channel Database because the vector of virtual sinks has all the necessary information about the environment to enable ray-tracing constrained channel estimation.

This channel estimation is based on the user spatial parameters estimation $(\hat{g}_{TX}, \hat{l}_{TX})$ and reconstructs the channel as,

$$\hat{h}_m = \hat{g}_{TX} \cdot \sum_{k=0}^{K-1} g_{m,k} \cdot Cvs(\hat{l}_{TX}, s_{m,k}) \cdot Str(\hat{l}_{TX}, l_{m,k}) \quad (1.11)$$

Because we have to estimate only 5 real parameters, user complex amplitude and location, much less than the $4K$ parameters required for multisource, the quality of such estimation will be higher.

- The antenna domain channel estimation has no constraints.
- The multibeam or multisource channel estimation reduces the channel estimation space to a subspace that satisfies the spatial constraints (final number of spatial beams/sources) and thereby improves estimation quality
- The multisink channel estimation reduces this subspace even more, to satisfy the spatial constraints and blockages of the specific environment described in the Channel Database and thereby improves estimation quality even more.

*B. Channel Estimation Algorithm*

Our channel estimation approach for multisink is as follows:
Antenna domain channel estimation:

$$\bar{h}_m \text{ for } m = 0, 1, \ldots M-1$$

- Find correlation peak (user location):

$$\hat{l}_{TX} = \arg\max_{l_{TX}} \left( \left| \sum_{m=0}^{M-1} \bar{h}_m \cdot \left( \sum_{k=0}^{K-1} g_{m,k} \cdot Cvs(l_{TX}, s_{m,k}) \cdot Str(l_{TX}, l_{m,k}) \right)^* \right| \right)$$
(1.12)

At the beginning we produce a search with rough resolution, $\lambda/4$ and then zoom in with more precise resolution. In our simulation we did a few iterations of zooming until we got resolution $\lambda/64$

- Given list of user location estimation we may estimate user complex amplitude according to:

$$\hat{g}_{TX} = \frac{1}{M} \cdot \sum_{m=0}^{M-1} \left( \sum_{m=0}^{M-1} \bar{h}_m \cdot \left( \sum_{k=0}^{K-1} g_{m,k} \cdot Cvs(\hat{l}_{TX}, s_{m,k}) \cdot Str(\hat{l}_{TX}, l_{m,k}) \right)^* \right)$$
(1.13)

- From spatial parameters list $(\hat{g}_k, \hat{l}_k)$ we reconstruct the channel according to (1.11):

*C. Simulation*

For multisink channel estimation simulation we use same simulation set up (see Figure 3) and same assumptions used for multisource channel estimation. We also assume the Channel Database that has full information about all mirror sinks of all AP antennas.

Because we know that the user is located within the room, we narrow our search space to from 0 to W and from 0 to D.

The simulation results for different numbers of antennas are shown in Figure 6, Figure 7 and Figure 8. It presents both multisource channel estimation EVM and multisink with data base assistance channel estimation EVM as function of antenna domain channel estimation EVM.

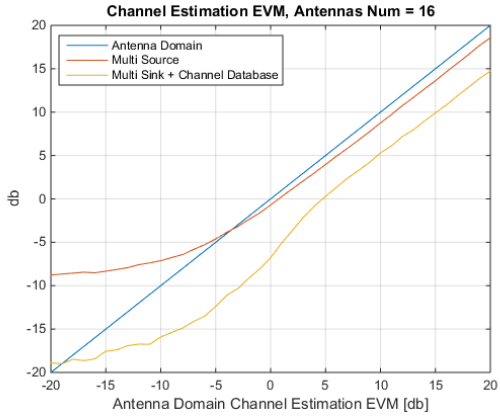

Figure 6: Channel Estimation EVM for 16 AP antennas.

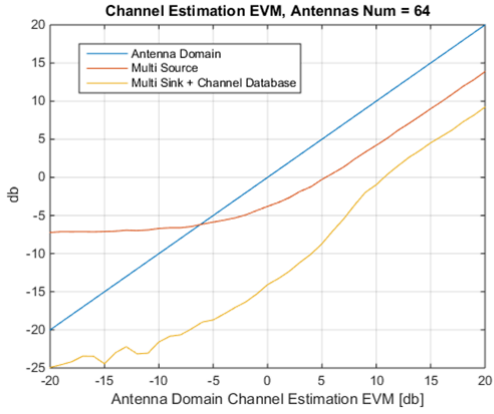

Figure 7 Channel Estimation EVM for 64 AP antennas.

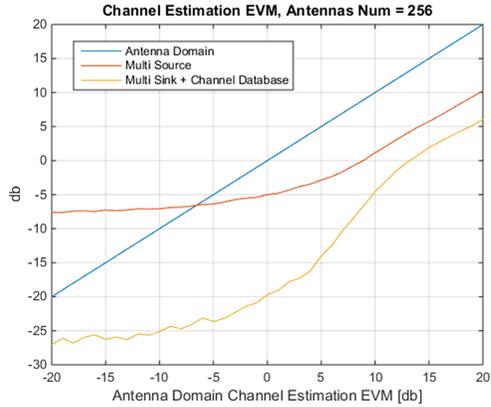

Figure 8 Channel Estimation EVM for 64 AP antennas.

IV. CHANNEL DATABASE CONCEPT

*A. Channel Database generation*

To generate the channel database we have to know the configuration of all surfaces in our environment. This information may be measured from visual data, from laser range-finders, or more interestingly, the AP can gradually build its Channel Database from our previous communication experience. Initially, the APs have no prior information, and may use the multisource approach to estimate the channel. The byproduct of this process is the list of virtual sources. By taking the symmetric line between the main and mirror sources we can deduce the location of the reflection surface, as is shown in figure below:

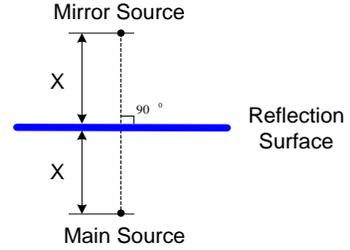

Figure 9: The Reflection Surface location determination

Comparing the gain of the main and virtual sources and accounting for the difference in distances, we can obtain the wall reflection coefficients.

The AP will gradually build up a 3D map of the environment, and may start to use the Channel Database approach. In real life, because there is no totally static environment, the location of virtual sinks will change with time, some new sinks will appear, others will disappear, etc. Moving objects will generate moving sinks and block known sinks. Therefore the multisink channel database approach has to be combined with the multisource approach. The channel database will provide information about the static part of the channel, while the multisource be will responsible for the dynamic part. It can discover new, unpredictable sources, track environment changes, and update the channel database.

*B. How practical is the Channel Database?*

We envision the channel database supporting a variety of communications and location algorithms that take advantage of *site-specific* information to increase capacity and improve location accuracy. However, its utility depends on how many channels may be accurately described with a limited number of virtual sinks (reflective surfaces). We believe that there are many spaces that may be described with relatively small number of flat surfaces. We call these 'low entropy' spaces. We may find such spaces in indoor or urban canyon scenarios. Other spaces will require an unrealistic number of reflective surfaces (virtual sinks). We call these spaces 'high entropy' spaces. An example of these spaces is shown in figure below:

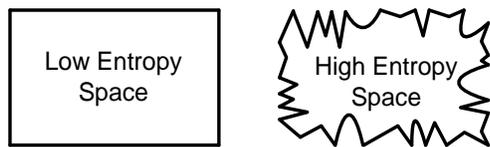

Figure 10: Low and high entropy spaces

The channel database assistance is an opportunistic approach that will work for low entropy spaces (indoor, urban canyon, etc.) The percentages of low vs. high entropy spaces that we will encounter in practice is a subject for future research.

## V. CONCLUSIONS

Applying geometric constraints from the local scattering environment leads to methods to improve channel estimation, based on an estimated position. We propose a channel estimation method that makes use of both 'virtual transmitters' (multisource) as well as 'virtual receivers' (multisink) to model multipath in low entropy environments. We plot the EVM of antenna domain channel estimation versus several forms of improved estimation. From our simulation results we draw the following conclusions:

Multisource turns out to have similar performance to beam domain channel estimation, however beam domain assumes a far-field description that does not apply to our scenario.

- When antenna domain channel estimation is very accurate, the multisource approach does not provide performance improvement and may even cause some degradation.
- However, when antenna domain channel estimation is poor, as typically happens with Massive MIMO in the low SNR region, multisource provides significant performance improvement,
- The larger the number of antennas, the larger the performance gain that multisource domain channel estimation provides, as large as 10 dB with 256 antennas

We also found that the channel database assistance provides significant performance gain, with typically a 4-5 dB gain using the multisink database over multisource. The larger the number of antennas, the larger the performance gain is.

This was only an initial performance evaluation to illustrate the potential, and can be applied to near-field as well as far-field. We envision the channel database supporting a variety of communications and location algorithms that take advantage of site-specific information to increase capacity and improve location accuracy.